\documentclass[10pt,aps,pre,twocolumn,superscriptaddress]{revtex4-1}
\usepackage{amsmath}
\usepackage{textcomp}
\usepackage{amssymb} 
\usepackage{graphicx}
\usepackage{tabularx}
\usepackage{placeins}
\usepackage{natbib}
\usepackage[normalem]{ulem}
\usepackage[usenames]{color}

\makeatletter
\newcommand*{\rom}[1]{\expandafter\@slowromancap\romannumeral #1@} 
\makeatother

\begin{document}

\title{Mechanical criticality of fiber networks at a finite temperature}

\author{Sadjad Arzash}
\altaffiliation[Present address: ]{Department of Physics, Syracuse University, Syracuse, NY}
\altaffiliation{Department of Physics \& Astronomy, University of Pennsylvania, Philadelphia, PA}
\affiliation{Department of Chemical \& Biomolecular Engineering, Rice University, Houston, TX 77005}
\affiliation{Center for Theoretical Biological Physics, Rice University, Houston, TX 77030}
\author{Anupama Gannavarapu}
\affiliation{Department of Chemical \& Biomolecular Engineering, Rice University, Houston, TX 77005}
\affiliation{Center for Theoretical Biological Physics, Rice University, Houston, TX 77030}
\author{Fred C.\ MacKintosh}
\affiliation{Department of Chemical \& Biomolecular Engineering, Rice University, Houston, TX 77005}
\affiliation{Center for Theoretical Biological Physics, Rice University, Houston, TX 77030}
\affiliation{Departments of Chemistry and Physics \& Astronomy, Rice University, Houston, TX 77005}

\begin{abstract}
At zero temperature, spring networks with connectivity below Maxwell's isostatic threshold undergo a mechanical phase transition from a floppy state at small strains to a rigid state for applied shear strain above a critical strain threshold. Disordered networks in the floppy mechanical regime can be stabilized by entropic effects at finite temperature. We develop a scaling theory for this mechanical phase transition at finite temperature, yielding relationships between various scaling exponents. Using Monte Carlo simulations, we verify these scaling relations and identify anomalous entropic elasticity with sub-linear $T$-dependence in the linear elastic regime. While our results are consistent with prior studies of phase behavior near the isostatic point, the present work also makes predictions relevant to the broad class of disordered thermal semiflexible polymer networks for which the connectivity generally lies far below the isostatic threshold.
\end{abstract}

\maketitle

\section{Introduction}

Fibrous materials are common in physiological systems that are responsible for the mechanical stability of cells and tissues. Examples include the interconnected network of biopolymers in the cytoskeleton and in the extracellular matrix. The linear elasticity of these biopolymer networks depends not only on the properties of the individual fibers but also on network architecture and specifically their connectivity, characterized by the local coordination number $z$. The key role of connectivity on the stability of mechanical structures has been well-established by Maxwell \cite{maxwell_l._1864} who showed that networks with Hookean, central-force (CF) interactions are linearly stable only when their average connectivity exceeds the isostatic threshold $z_c = 2d$, where $d$ is dimensionality. For physiological networks, however, this rigidity transition is not relevant, as their connectivity lies well below this threshold \cite{lindstrom_biopolymer_2010,lindstrom_finite-strain_2013,jansen_role_2018} and network stability depends on non-CF interactions such as fiber bending rigidity \cite{satcher_theoretical_1996,kroy_force-extension_1996,head_deformation_2003,wilhelm_elasticity_2003,das_effective_2007,wyart_elasticity_2008,zaccone_approximate_2011,broedersz_criticality_2011}. 
Recent theory and experimental studies have identified a strain-controlled rigidity transition for networks of fibers such as collagen, e.g., for shear strains above a critical threshold $\gamma_c$ \cite{sharma_strain-controlled_2016}. Moreover, this transition exhibits rich critical phenomena, including scaling behavior and non-mean-field effects \cite{rens_nonlinear_2016,feng_nonlinear_2016,vermeulen_geometry_2017,rens_micromechanical_2018,shivers_scaling_2019,merkel_minimal-length_2019,arzash_finite_2020,arzash_shear-induced_2021,damavandi_energetic_2022,lerner_anomalous_2023}.  
But, these prior studies of fiber systems have been limited to athermal networks and little is know of the effects of thermal fluctuations that can be expected to stabilize mechanically floppy systems and lead to entropic elasticity \cite{de_gennes_scaling_1979,plischke_entropic_1998,plischke_entropic_1999,farago_entropic_2000}. Prior simulations and mean-field theory have pointed to critical signatures for the isostatic transition at finite temperature $T$ \cite{dennison_fluctuation-stabilized_2013,zhang_finite-temperature_2016, dennison_critical_2016}. Additionally, in a study on random-bond sub-isostatic networks under finite isotropic strain, Ref.\ \cite{wigbers_stability_2015} found that thermal fluctuations stabilize a floppy network with an anomalous temperature dependence near a critical bulk strain that appears to coincide with what was found at the isostatic point. But, a theory for the mechanical criticality and the corresponding exponents remain unclear for the strain-controlled transition in sub-isostatic systems at a finite temperature, including the broad class of semiflexible polymers.

\begin{figure}[!h]
	\centering
	\includegraphics[width=8.5cm, height=8.5cm,keepaspectratio]{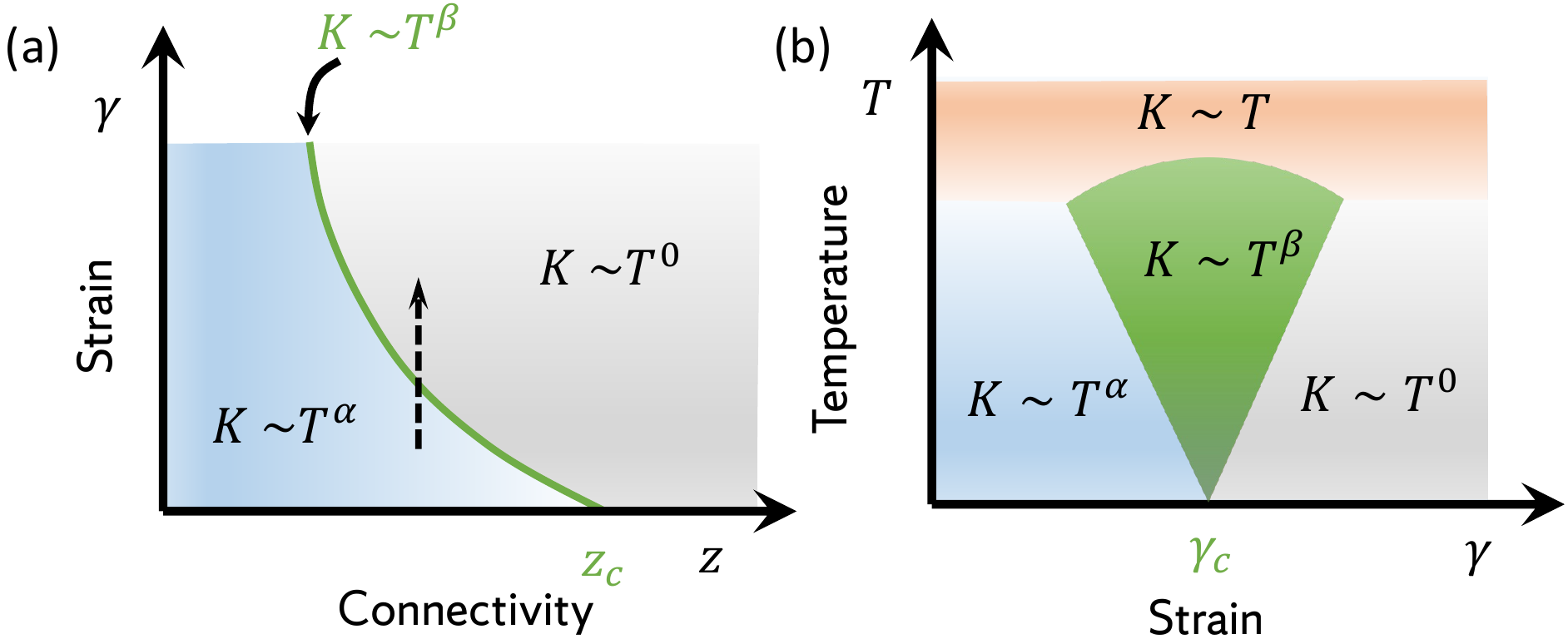}
	\caption{\label{fig_phase_diagram} 
	Schematic phase diagrams of disordered spring networks in the limit of low temperature $T$ (a) and finite $T$ (b). The shear stiffness $K$ exhibits different scaling behavior with temperature based on the network's connectivity $z$ and the applied shear strain $\gamma$. In the limit of small $\gamma$, $K$ reduces to the linear shear modulus. (a) With increasing strain, mechanically floppy (subisostatic) networks with $z<z_c$ crossover from entropic to enthalpic, stretching-dominated behavior in the vicinity of the $T=0$ phase boundary (dashed arrow). (b) With increasing $T$, critical behavior extends to a broad zone about $\gamma_c$ in which the $T$-dependence changes. }
\end{figure}

Here, we study the critical behavior of the strain-controlled rigidity transition at finite temperature by performing Monte Carlo (MC) simulations of central-force spring networks. In the linear regime, we find an anomalous entropic elastic regime that extends throughout the regime with $\gamma<\gamma_c$ in Fig.\ \ref{fig_phase_diagram}b. Here, the linear shear modulus varies with $T$ as $G\sim T^\alpha$, with an exponent $\alpha\simeq0.8$. Along the line at $\gamma=0$, these results are consistent with Ref.\ \cite{dennison_fluctuation-stabilized_2013}. For shear strains $\gamma > \gamma_c(z)$, the network's elastic response becomes independent of temperature, consistent with the stretching-dominated regime previously seen for connectivities $z>z_c$. We also develop a scaling theory that not only provides a theoretical framework for these results but also allows us to identify scaling relations among various critical exponents, which we also test here. We also quantify the network's fluctuations that can have either thermal or athermal, nonaffine origin. We find a peak in the fluctuations near the critical strain, analogous to prior results for athermal systems. In contrast to temperature controlled phase transitions, temperature $T$ acts as a stabilization effect or field and moves the system away from criticality, analogous to quantum critical points at zero temperature \cite{sachdev_quantum_2011}. Similar to such systems, we also find that the effects of criticality extend to finite temperature as illustrated in Fig.\ \ref{fig_phase_diagram}a all along the critical line given by $\gamma_c(z)$.

\section{Scaling theory}

The nonlinear mechanics of fiber networks at zero temperature has been explained in terms of a bending-dominated to a stretching-dominated critical transition that occurs at a critical shear strain $\gamma_c(z)$, which depends on the network's connectivity $z$ and architecture \cite{sharma_strain-controlled_2016,sharma_strain-driven_2016,licup_elastic_2016,rens_nonlinear_2016,feng_nonlinear_2016}. 
We develop a scaling theory inspired by real-space renormalization arguments introduced by Kadanoff \cite{kadanoff_scaling_1966}. The critical signatures of this strain-controlled mechanical phase transition have been recently examined using a real-space renormalization approach and finite-size scaling methods and recently extended to athermal networks \cite{shivers_scaling_2019,arzash_finite_2020,arzash_shear-induced_2021}. 
For finite temperature $T$, however, we consider the system's free energy $F$ per network element, e.g., mesh or strand. As with other critical phenomena, we focus on the \emph{singular} part $F$ as a function of reduced strain $t=\gamma-\gamma_c$ and $T$, noting that strain $\gamma$ is the control variable for the transition at $t=0$ and $T$ is an auxiliary field that moves the system away from the (athermal) critical point. We expect critical signatures such as fluctuations and singularities as both $t$ and $T\rightarrow0$. 

Under rescaling of the system by a factor $L$, we expect the system to exhibit a homogenous free energy density near criticality, for which 
\begin{equation}\label{eq_scaled_free_energy}
	F(t,T) = L^{-d} F(t L^x, T L^y),
\end{equation}
where $d$ is the dimensionality and $x,y>0$ are fundamental exponents. 
The mechanical quantities such as shear stress $\sigma$ and the shear stiffness or differential shear modulus $K$ are obtained by taking the first and second derivatives of $F$ with respect to strain, i.e., $t$. Thus, 
\begin{equation}\label{eq_modulus_scaling_relation}
	K = \frac{\partial \sigma}{\partial \gamma} \sim \frac{\partial^2 F(t,T)}{\partial t^2} \sim L^{-d+2x}F_{2,0}(tL^x, TL^y)
\end{equation}
where $F_{n,m}$ refers to the $n$th partial derivative with respect to $t$ and $m$th partial derivative with respect to $T$ of $F$. Since the rescaling factor $L$ is an arbitrary parameter, we can substitute $L=|t|^{-1/x}$ in Eq.\ \eqref{eq_modulus_scaling_relation}. This identifies the correlation length exponent $\nu=1/x$ and leads to a scaling function
\begin{equation}\label{eq_mc_widom}
	K = |\gamma - \gamma_c|^f \mathcal{G}_{\pm}(T/|\gamma - \gamma_c|^\psi),
\end{equation}
where $f = d\nu - 2$ and $\psi = y\nu$. Moreover, to ensure the continuity of function $F_{2,0}(\pm1,s)$ at the critical point $t \rightarrow 0$, we must have $F_{2,0}(\pm1,s) \sim s^{f/\psi}$. This power law relation provides the $T-$dependence behavior of $K$ at $\gamma_c$, i.e., $K(\gamma_c) \sim T^{\beta}$, where
	$\beta = f/\psi$.

\section{Model}

In order to study the effects of temperature in fiber networks, we perform Monte Carlo simulations in 2D systems using the triangular network model. Starting from a full triangular network with $z=6$, we randomly cut bonds until a desired subisostatic connectivity $z<z_c$ is reached. We remove the dangling nodes since they have no mechanical contribution to the network's response. Here, we simulate networks at an average connectivity of $z=3.3$. A small section of such model is shown in Appendix A.

The network's elastic energy is limited to central force interactions only, i.e., there is no bending energy in our models. The energy is given by
\begin{equation}\label{eq_mc_energy}
	E = \frac{\mu}{2} \sum_{ \langle ij \rangle}^{}\frac{ (l_{ij} - l_{ij,0}) ^2 }{l_{ij,0}},
\end{equation}
where $l_{ij,0}$ and $l_{ij}$ are the initial and current bond length between nodes $i$ and $j$, respectively, and $\mu$ is the stretching stiffness of the bonds. The summation is over all nodes in the network. We note that in order to isolate the influence of thermal fluctuations, we have not included bending interactions in our model. Incorporating a finite bending rigidity $\kappa$ would require an additional scaling variable $\kappa/|\gamma-\gamma_c|^\phi$ in Eq. (3). There would also be a competition between these two stabilizing effects. Nevertheless, our results for $\kappa=0$ can be expected to represent a good approximation for weakly bending systems, particularly such as intermediate filament or fine-clot fibrin networks, where their thermal persistence length is comparable to the mesh size
\cite{broedersz_modeling_2014, lin_origins_2010,kpiechocka_multi-scale_2016}. Describing the full temperature and bending dependence, however, will become a challenging task.

We simulate a system with $N$ nodes in a volume $V$ using MC simulations in the canonical ensemble. We set the stretching stiffness $\mu=1$ and vary the reduced temperature $T \equiv k_BT/\mu l_c^2$, where $k_{B}$ is the Boltzmann constant and $l_c = \langle l_{ij,0} \rangle$ is the average initial bond length in networks, which is $1.0$ in our model. After applying a shear strain, we find the minimum energy configuration at zero temperature using FIRE \cite{bitzek_structural_2006} and let the system reach its equilibrium configuration at $T$ by running at least $\tau_{\textrm{eq}} = 10^7$ MC steps with a trial move size chosen to yield a $50\%$ acceptance ratio based on the Metropolis algorithm \cite{metropolis_equation_1953,frenkel_understanding_2002}. We calculate average energy and stress components using simulations over $\tau_{\textrm{run}} = 10 \tau_{\textrm{eq}}$ MC steps. The stiffness $K$ is obtained from the average shear stress $\sigma$ as $K = \partial \sigma/\partial \gamma$ \cite{dennison_fluctuation-stabilized_2013} (see Appendix A). The data are an ensemble average of 10 random samples, unless otherwise stated.

\section{Results}

\begin{figure}[!h]
	\centering
	\includegraphics[width=6cm, height=6cm,keepaspectratio]{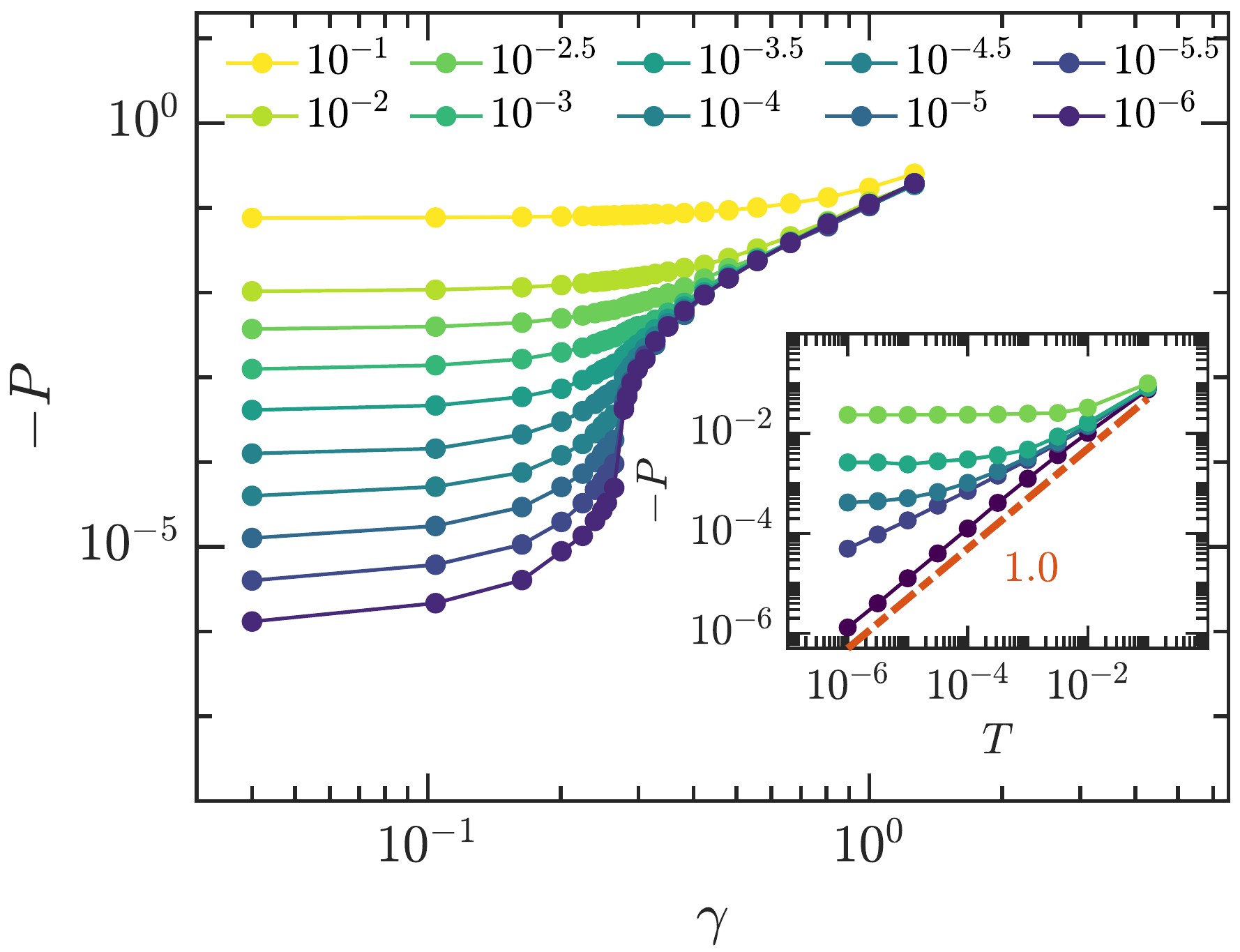}
	\caption{\label{fig_mc_p} Pressure as a function of shear strain for diluted triangular networks with $z=3.3$ and varying temperature. The lateral system size is $W=50$. The inset shows the behavior of $P$ versus $T$ at five different strain values. The lowest data points are in the linear regime ($\gamma<\gamma_c$), the next one is at $\gamma_c$. The upper three data sets are for large strains where $\gamma>\gamma_c$.}
\end{figure}

We first study the behavior of internal pressure $P$ (see Appendix A) for thermal networks as a function of shear strain and temperature. As shown in Fig.\ \ref{fig_mc_p}, we find that thermal networks are under tension, i.e., $P<0$. As we increase $\gamma$, the potential energy between the nodes increases, which results in a larger absolute value of $P$ (Fig.\ \ref{fig_mc_p}). The dependence of pressure versus temperature is shown in the inset of Figure\ \ref{fig_mc_p} for five different values of $\gamma$. In the linear regime where $\gamma<\gamma_c$, we find that the magnitude of $P$ is linearly increasing with $T$, i.e., the system's pressure is dominated by the ideal gas effects. This is in agreement with Ref.\ \cite{dennison_fluctuation-stabilized_2013}. As we increase $\gamma$ close to $\gamma_c$, however, the $T$-dependence of $P$ becomes sublinear. At very large strains, pressure has no temperature dependence (inset of Fig.\ \ref{fig_mc_p}).

\begin{figure}[!h]
	\centering
	\includegraphics[width=8cm, height=8cm,keepaspectratio]{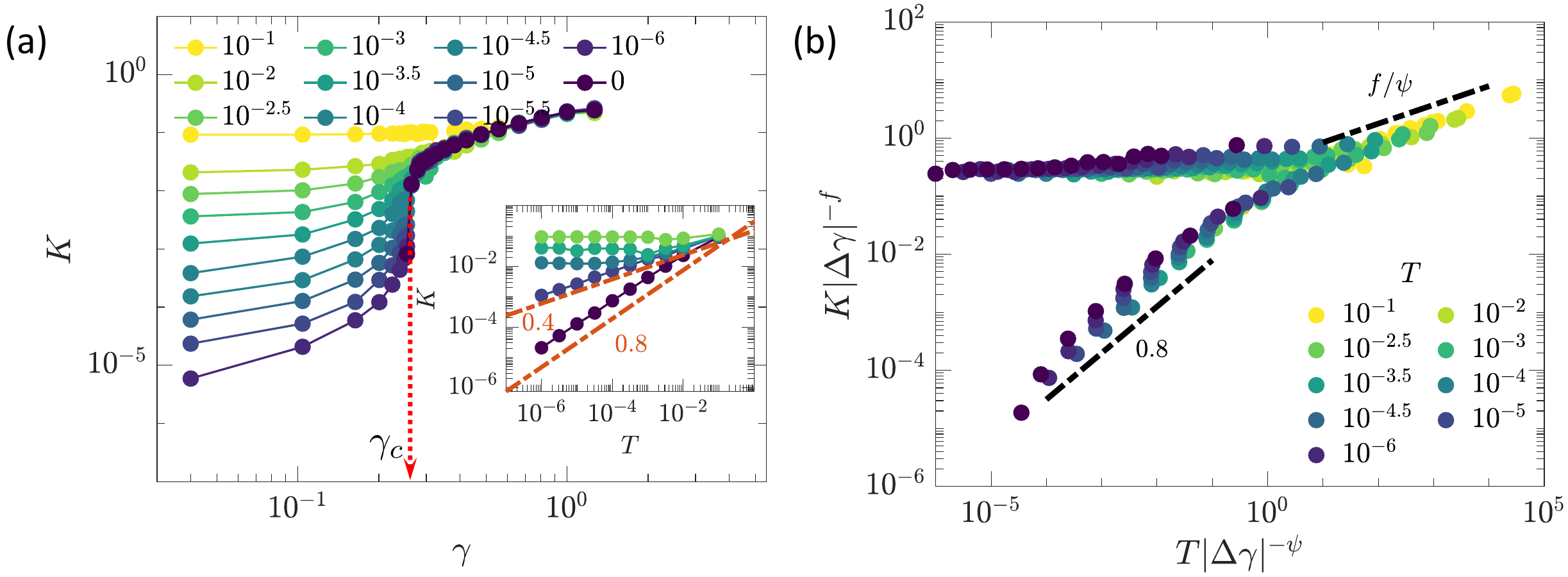}
	\caption{\label{fig_mc_widom} (a) Shear stiffness or differential shear modulus of diluted triangular networks at $z=3.3$ versus strain for various temperatures as indicated in the legend. The system size is $W=50$ here. The critical strain is indicated by the red arrow. Inset: shear modulus versus temperature at four different shear strains: the lowest curve is in the linear regime where $\gamma<\gamma_c$, the second curve is at $\gamma_c$. The upper two data sets are for large strains where $\gamma>\gamma_c$. (b) The Widom-like collapse of the data in (a) using the critical exponents $f = 0.76$ and $\psi = 2.35$.}
\end{figure}

Figure\ \ref{fig_mc_widom}a shows the shear stiffness $K$ as a function of shear strain $\gamma$ for various reduced temperatures $T$. In the small strain regime, as $T$ increases, the network stiffness increases with an anomalous $T$-dependence exponent of $0.8$. This anomalous entropic elasticity is consistent with prior results for the linear shear modulus \cite{dennison_fluctuation-stabilized_2013}, although we observe this throughout the (central-force) floppy region indicated by blue in Fig.\ \ref{fig_phase_diagram}a. For strains beyond $\gamma_c$, the network's response becomes independent of temperature because of highly stretched bonds. This mechanical response depends only on the network structure and strain magnitude. At the critical strain, on the other hand, we find that the stiffness exhibits a different anomalous scaling behavior with $K \sim T^{0.4}$ (inset of Fig.\ \ref{fig_mc_widom}a)) similar to prior results and mean-field predictions at the isostatic point at $z_c$ and $\gamma=0$ in Fig.\ \ref{fig_phase_diagram}a\cite{dennison_fluctuation-stabilized_2013,zhang_finite-temperature_2016}. By estimating the critical exponents $f$ and $\psi$, we collapse the modulus data in Fig.\ \ref{fig_mc_widom}b according to the scaling function in Eq. \eqref{eq_mc_widom}. The exponent $f$ is found from the supercritical regime $\gamma > \gamma_c$ at zero temperature, where $K - K_c \sim |\gamma - \gamma_c|^f$. For this system size, we find $f = 0.76 \pm 0.14$. We select the value of exponent $\psi$ that leads to the optimal collapse of our data. The apparent deviation observed in the regime close to $\gamma_c$ of this collapse is related to the finite size effects in our simulations; if the correlation length becomes comparable or larger than the system size, which can occur for strains close to the critical point, then the simulations are incapable of capturing the critical effects \cite{arzash_finite_2020}. These exponents are in good agreement with our derived relation $K(\gamma_c) \sim T^{f/\psi}$.

\begin{figure}[!h]
	\centering
	\includegraphics[width=8cm, height=8cm,keepaspectratio]{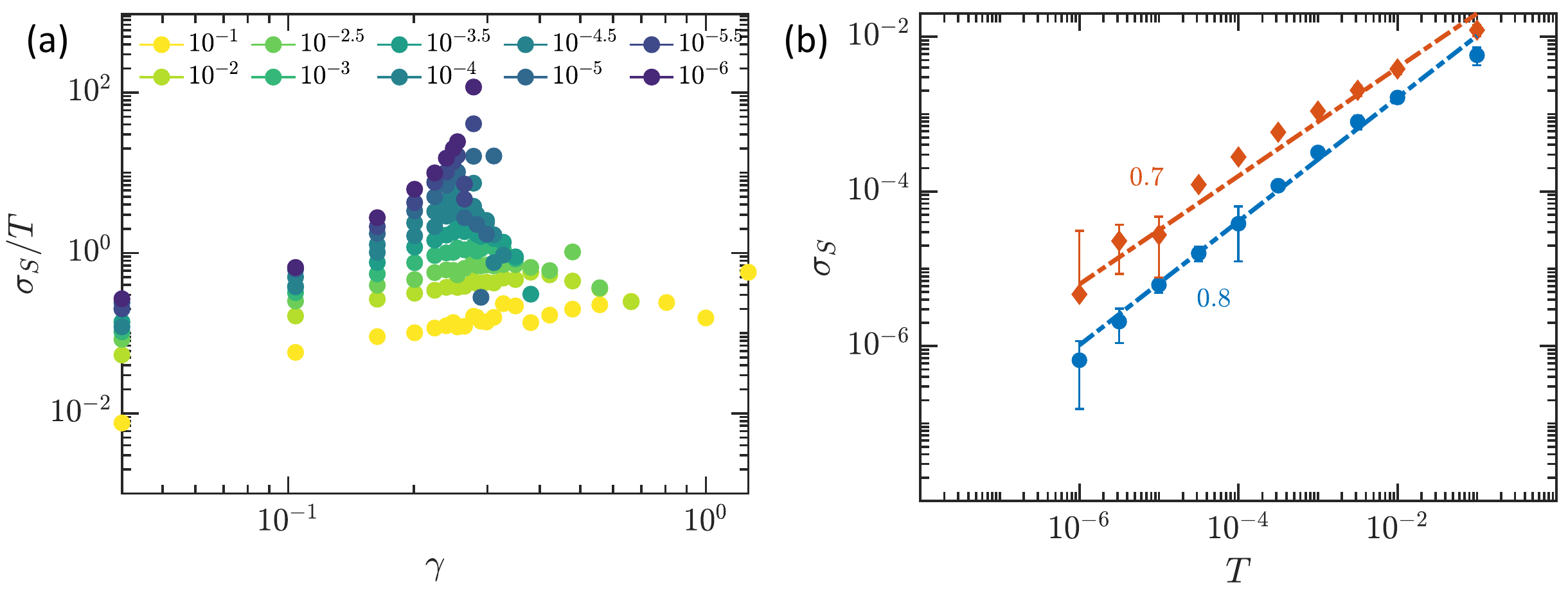}
	\caption{\label{fig_mc_sigma_s} (a) The entropic stress $\sigma_S$ scaled with temperature versus shear strain in diluted triangular networks at $z=3.3$. (b) The scaling behavior of $\sigma_{S}$ versus temperature in the linear regime (blue circles) and at the critical strain (red diamonds).}
\end{figure}

To explore entropic elasticity in these thermal networks, it is informative to identify the entropic contribution to the stress and its scaling behavior. 
We extend the scaling theory above to identify entropic effects more directly by taking derivatives with respect to $T$. Denoting the system's entropy as $S$ and noting that $F = E - T S$, we can divide the stress contributions in two parts as
\begin{equation}\label{eq_stress_contributions}
	\sigma = \frac{1}{V} \frac{\partial F}{\partial \gamma} = \frac{1}{V} \frac{\partial E}{\partial \gamma} - \frac{T}{V} \frac{\partial S}{\partial \gamma},
\end{equation}
where the first term is the enthalpic contribution $\sigma_{E}$ and the second term is the entropic part $\sigma_{S}$. In a canonical ensemble, we have $S = - (\partial/\partial T)F$ \cite{mcquarrie_statistical_1975}. 
Classic entropic elasticity is characterized by stress and moduli that scale linearly with $T$, e.g., for which $\sigma=\sigma_{S}$ and one should observe a $T$-independent behavior of $\sigma_{S}/T$. 
In our diluted disordered networks, however, $\sigma_{S}/T$ \cite{dennison_fluctuation-stabilized_2013} shows a strong dependence on temperature (Fig.\ \ref{fig_mc_sigma_s}a). As we approach the critical strain, $\sigma_{S}/T$ exhibits a diverging behavior at low temperature. 
This is consistent with our scaling theory, where the entropic stress is expected to behave as
\begin{equation}\label{eq_sigma_s_scaling}
	\sigma_{S} = T |t|^{f- \psi + 1} F_{1,1}(\pm1, T/|t|^\psi).
\end{equation}
At $\gamma=\gamma_c$, continuity of this requires that 
$\sigma_{S} \sim T^{({f+1})/{\psi}}$. Together with the previously identified exponents $f\simeq0.76$ and $\psi\simeq2.35$, this prediction can account for the anomalous $T-$dependence near the critical point in Fig.\ \ref{fig_mc_sigma_s}a and b. Specifically, at the critical strain, our simulations revealed that $\sigma_{S}$ follows a power law scaling with temperature, with an exponent of 0.7. 
In the linear regime, however, the entropic contribution is dominant (see Appendix B) and $\sigma=\sigma_S$ and $K \sim  T^\alpha$, where 
$\mathcal{G}_{-}(s)\sim s^\alpha$ (Fig.\ \ref{fig_phase_diagram}a), which is also consistent with what we observe in Fig.\ \ref{fig_mc_sigma_s}b. 

The anomalous temperature dependence of shear modulus in diluted networks is due to disorder. MC simulations on 1D chains of springs show expected entropic elasticity \cite{dennison_fluctuation-stabilized_2013}. To gain insight into the sublinear dependence of $K$ on temperature at the critical strain, we chose to investigate the honeycomb lattice model without any distortion (see Appendix C). Due to its symmetry, this model exhibits $\gamma_c = 0.0$ \cite{rens_nonlinear_2016}, i.e., honeycomb lattice is critically stable in the linear regime. We find that $K \sim T^{0.5}$ for this model in the linear regime, which resembles the behavior of shear modulus in diluted triangular networks near their critical connectivity \cite{dennison_fluctuation-stabilized_2013,zhang_finite-temperature_2016}.

\begin{figure}[!h]
	\centering
	\includegraphics[width=8cm, height=8cm,keepaspectratio]{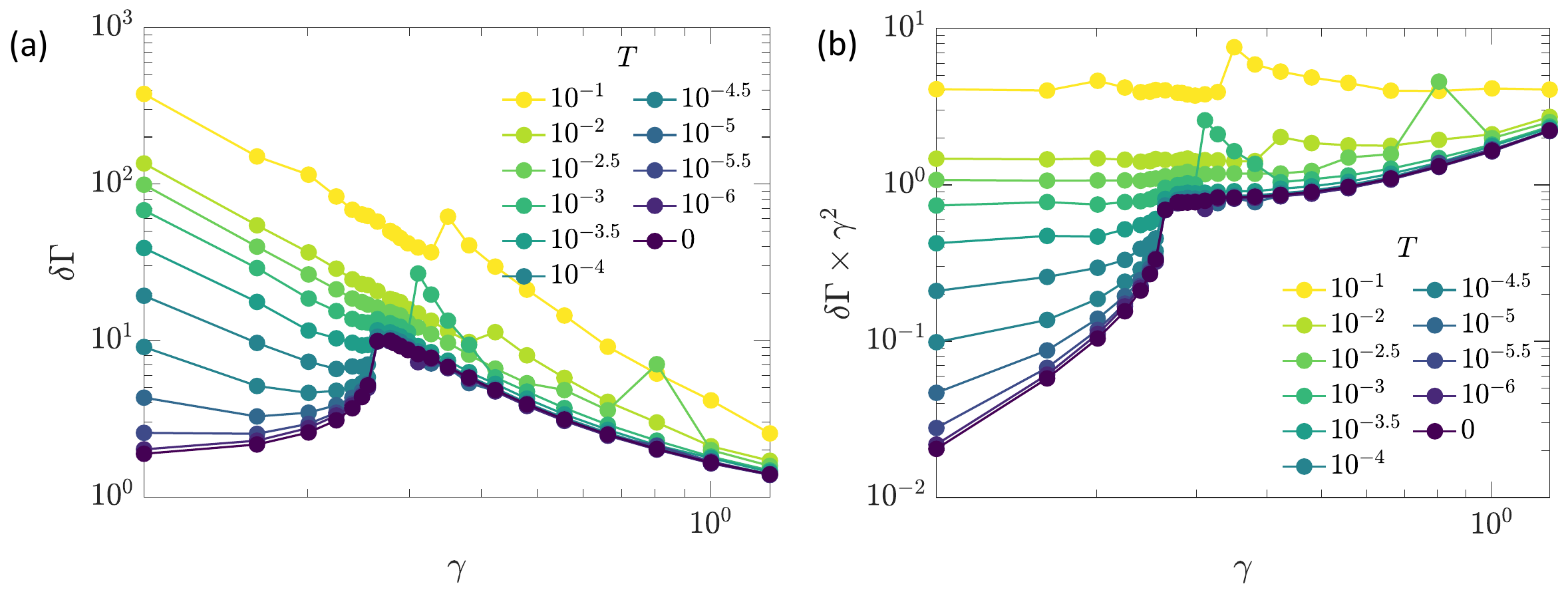}
	\caption{\label{fig_mc_fluctuations} (a) The fluctuations calculated from Eq.\ \eqref{eq_mc_fluctuations} as a function of strain in diluted triangular networks with $z=3.3$ and varying temperature. At $T=0$, these nonaffine fluctuations exhibit a peak at the critical strain. As the temperature increases, the mechanical criticality of the system becomes less pronounced. (b) The same data in (a) that are scaled with the applied strain magnitude.}
\end{figure}

One of the most striking features of a critical phase transition is the divergence of fluctuations near the critical point. Following Ref.\ \cite{tauber_role_2020}, we calculate these fluctuations in our thermal networks as
\begin{equation}\label{eq_mc_fluctuations}
	\delta \Gamma = \frac{\overline{\langle (\mathbf{u} - \overline{\mathbf{u}}_{\textrm{aff}})^2 \rangle}}{\ell_c^2 \delta \gamma^2},
\end{equation}
where the bars indicate MC averages and the angular brackets represent the averages over nodes and random samples, $\ell_c$ is the average initial position of the bonds (which is 1.0 in lattice models), $\delta \gamma$ is the imposed strain step, $\overline{\mathbf{u}}_{\textrm{aff}}$ is the affine location of the node's position that was obtained using the MC averages of the previous strain step, and $\mathbf{u}$ is the instantaneous position of the node during current MC simulation run. For low values of $T$, $\delta \Gamma$ exhibits a peak at the critical strain (Fig.\ \ref{fig_mc_fluctuations}). At high temperatures, however, the system moves further from criticality and the large thermal fluctuations suppress the critical effects in this strain-controlled transition, thus, the peak vanishes. For $T=0$, the fluctuations are suppressed, as expected for finite-size effects. We also note that the apparent scaling behavior of $\delta \Gamma \sim \gamma ^{-2}$ away from $\gamma_c$ is a trivial effect of our definition in Eq.\ \eqref{eq_mc_fluctuations} (This is shown in Fig.\ \ref{fig_mc_fluctuations}b). By examining the fluctuations near $\gamma_c$, we confirm that the finite temperature effects smear out the criticality in these disordered systems, analogous to zero-temperature criticality in quantum systems \cite{sachdev_quantum_2011,dennison_fluctuation-stabilized_2013}.

\section{Conclusions}

Our results show that thermal fluctuations can stabilize mechanically floppy networks in a way similar to the addition of bending or other interactions.  We also find anomalous entropic elasticity with a corresponding exponent $\alpha\simeq0.8$ of $T$ throughout the regime of strains $\gamma<\gamma_c$. This is quantitatively consistent with prior simulations of the linear (small strain) regime \cite{dennison_fluctuation-stabilized_2013} and qualitatively consistent with a prior mean-field theory for which the (mean-field) exponent $\alpha=1$ \cite{zhang_finite-temperature_2016}. This anomalous entropic elasticity with exponent $\alpha\simeq0.8$ is, however, only expected for systems that are sufficiently far from criticality. In the vicinity of the critical line in Fig.\ \ref{fig_phase_diagram}a, a smaller exponent $\beta$ close to 1/2 is observed. This is similar to what has been reported near the isostatic point \cite{dennison_fluctuation-stabilized_2013,zhang_finite-temperature_2016}. Ref.\ \cite{lee_generic_2023} also reports an exponent close to 1/2, consistent with the near critical behavior in Refs.\ \cite{dennison_fluctuation-stabilized_2013, zhang_finite-temperature_2016} and Fig. 3a above. We note that the authors of Ref.\ \cite{lee_generic_2023} combine bulk strain with shear, resulting in a very small floppy regime (blue in Fig. 1) that likely makes identification of the exponent $\alpha$ difficult.


The observed anomalous temperature dependence is closely related to the behavior of entropic contributions. We find that the entropic stress $\sigma_{S}$ dominates the response in the linear regime albeit with a sublinear $T$-dependence. However, our results also suggest that such singular signatures of criticality associated with the transition in Fig.\ \ref{fig_phase_diagram} may be dominated by non-singular thermal effects such as for the pressure in Fig.\ \ref{fig_mc_p}. Fundamentally, shear stress is insensitive to ideal gas-like contributions arising from thermal fluctuations. Thus, in order to test these predictions experimentally, it will be important to focus on volume-preserving simple shear, as is the case with most rheometers \cite{HeatCapacity}.

Although we have focused on networks of Hookean springs, our results should also apply to the broad class of semiflexible polymer networks such as those of cytoskeletal polymers \cite{carrillo_nonlinear_2013,broedersz_modeling_2014,meng_theory_2017} or related synthetic networks \cite{kouwer_responsive_2013,jaspers_ultra-responsive_2014}, although whether bending or thermal effects dominate can be expected to depend on the thermal persistence length $\ell_p$ and network mesh
size, as discussed above. It would also be interesting to explore whether other non-thermal fluctuation phenomena, such as active stress fluctuations due to molecular motors in cytoskeletal networks \cite{mizuno_nonequilibrium_2007,koenderink_active_2009,winer_non-linear_2009,jansen_cells_2013,sheinman_actively_2012} may also lead to qualitatively similar fluctuation stabilization and possibly even a phase diagram similar to Fig.\ \ref{fig_phase_diagram}a.

\section*{Acknowledgments}

This work was supported in part by the National Science Foundation Division of Materials Research (Grant No. DMR-2224030) and the National Science Foundation Center for Theoretical Biological Physics (Grant No. PHY-2019745). We also would like to acknowledge our insightful discussions with Tom Lubensky.


\appendix
\renewcommand\thefigure{A\arabic{figure}} 
\setcounter{figure}{0}
\renewcommand\theequation{A\arabic{equation}}
\setcounter{equation}{0}


\section{Details of MC simulations}

Because we aim to explore thermal fluctuations as a stabilization effect, the network's elastic energy is limited to central force interactions only, i.e., there is no bending energy in our models. The energy is given by
\begin{equation}\label{eq_mc_energy}
	E = \frac{\mu}{2} \sum_{ \langle ij \rangle}^{}\frac{ (l_{ij} - l_{ij,0}) ^2 }{l_{ij,0}},
\end{equation}
where $l_{ij,0}$ and $l_{ij}$ are the initial and current bond length between nodes $i$ and $j$, respectively, and $\mu$ is the stretching stiffness of the bonds. The summation is over all nodes in the network. We note that there is no non-bonded interactions such as excluded volume effects in our model, i.e., the springs can potentially overlap during simulation. The macroscopic volume-preserving shear strain $\gamma$ is applied in the $x-$direction using the following deformation tensor
\begin{equation} \label{eq_mc_shear_tensor}
	\Lambda(\gamma) = \begin{bmatrix}
		1 & \gamma \\
		0 & 1
	\end{bmatrix},
\end{equation}
To minimize the edge effects, we use periodic boundary conditions in all directions. Furthermore, we utilize Lees-Edwards boundary conditions \cite{lees_computer_1972} in order to shear our systems. The stress components are calculated as following \cite{doi_theory_1988}
\begin{equation} \label{eq_mc_stress}
	\sigma_{\alpha \beta} = \frac{1}{2V} \sum_{ij}^{}f_{ij,\alpha} r_{ij,\beta}
\end{equation}
where $V$ is the volume (area) of the system, $f_{ij,\alpha}$ is the $\alpha$ component of the force exerted on node $i$ by node $j$, and $r_{ij,\beta}$ is the $\beta$ component of the displacement vector connecting nodes $i$ and $j$. The summation is taken over all nodes in the network.

At every shear strain $\gamma$ for a network at temperature $T$, we calculate the pressure as
\begin{equation}\label{eq_mc_p}
	P = \frac{NT}{V} - \frac{1}{d} (\sum_{i} \sigma_{ii}),
\end{equation}
where $N$ is the number of nodes, $T$ is temperature in reduced units, $V$ is the volume of the system, $d$ is dimensionality, and $\sigma_{ii}$ are the normal components of the stress tensor in Eq. \eqref{eq_mc_stress} that are averaged over MC simulations. The first term in this equation is due to the ideal gas contributions of the nodes and the second part comes from the potential interactions.

For a system with $N$ nodes in a volume $V$, the MC simulations are performed in the canonical ($NVT$) ensemble. We fix the stretching stiffness $\mu=1$ in our simulations and vary the reduced temperature $T \equiv k_BT/\mu l_c^2$, where $k_{B}$ is the Boltzmann constant and $l_c = \langle l_{ij,0} \rangle$ is the average initial bond length in networks, which is $1.0$ in our triangular lattice. After applying a shear strain, we use FIRE \cite{bitzek_structural_2006} to find the minimum energy configuration at zero temperature. We then let the system reach equilibrium at the desired temperature by performing MC steps.

We use the standard Metropolis algorithm \cite{metropolis_equation_1953,frenkel_understanding_2002} to perform MC moves. At every MC step, we randomly displace all nodes in our elastic network with a magnitude $\delta$. The move is accepted with a probability $\textrm{min}\big(1, \exp( -\beta[ E_{\textrm{new}}  - E_{\textrm{old}} ] )  \big)$, where $\beta = 1/T$ (note that $T \equiv k_BT/\mu l_c^2$), $E_{\textrm{old}}$ and $E_{\textrm{new}}$ are the energy values calculated from Eq. \eqref{eq_mc_energy} in the main text before and after the MC move, respectively. The parameter $\delta$ plays a crucial role in the efficiency of sampling the configuration space. If the value of $\delta$ is too large, the newly generated configurations are likely to have very high energy, and hence, the trial move will likely be rejected. Conversely, if the value of $\delta$ is too small, the newly generated configurations will have a similar elastic energy to the previous configuration, and most moves will be accepted. Therefore, finding an appropriate value for $\delta$ is essential for efficient sampling of the configuration space. During the simulations, we track the acceptance ratio of Monte Carlo (MC) moves and dynamically adjust the value of $\delta$ to maintain an acceptance ratio of around 50\%.

\begin{figure}[!h]
	\centering
	\includegraphics[width=8.5cm, height=8.5cm,keepaspectratio]{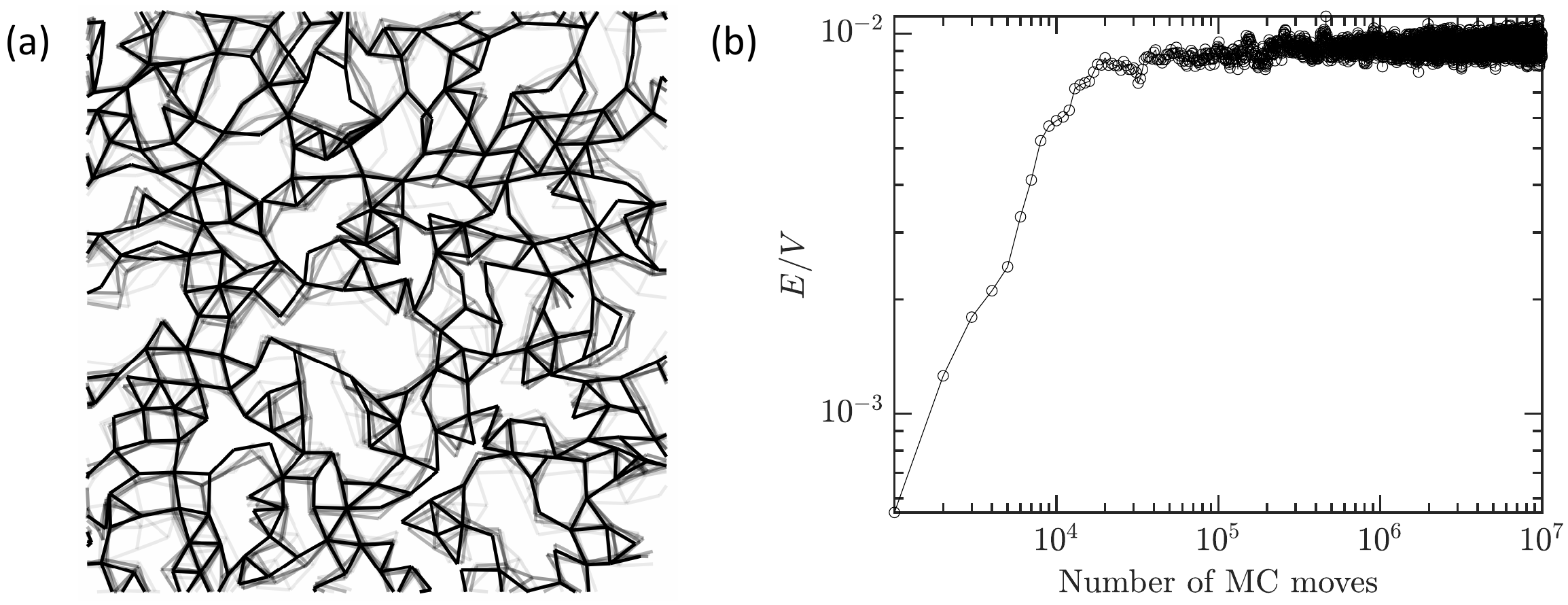}
	\caption{\label{fig_sup_mc_equilibration}(a) Showing multiple snapshots (corresponding to different shades of gray) of MC configurations of a diluted triangular network under a shear strain $\gamma < \gamma_c$ at a finite $T = 10^{-2}$ during equilibration step. (b) The instantaneous value of elastic energy density versus MC moves at $T = 10^{-2}$ for a diluted triangular model at $z=3.3$ and under a shear strain $\gamma<\gamma_c$.}
\end{figure}

To ensure that the system is in equilibrium, we track the elastic energy and shear stress as a function of the number of Monte Carlo moves. We have determined that our models are well equilibrated after running MC for $\tau_{\textrm{eq}} = 10^7$ moves (see Fig. \ref{fig_sup_mc_equilibration}b). After this equilibration step, we perform ensemble averaging for a duration of $\tau_{\textrm{run}} = 10^8$ MC moves. Figure \ref{fig_sup_mc_equilibration}a shows some snapshots of our MC simulations.

\renewcommand\thefigure{B\arabic{figure}} 
\setcounter{figure}{0}
\renewcommand\theequation{B\arabic{equation}}
\setcounter{equation}{0}

\section{Shear stress behavior of diluted triangular networks}

\begin{figure}[!h]
	\centering
	\includegraphics[width=5cm, height=5cm,keepaspectratio]{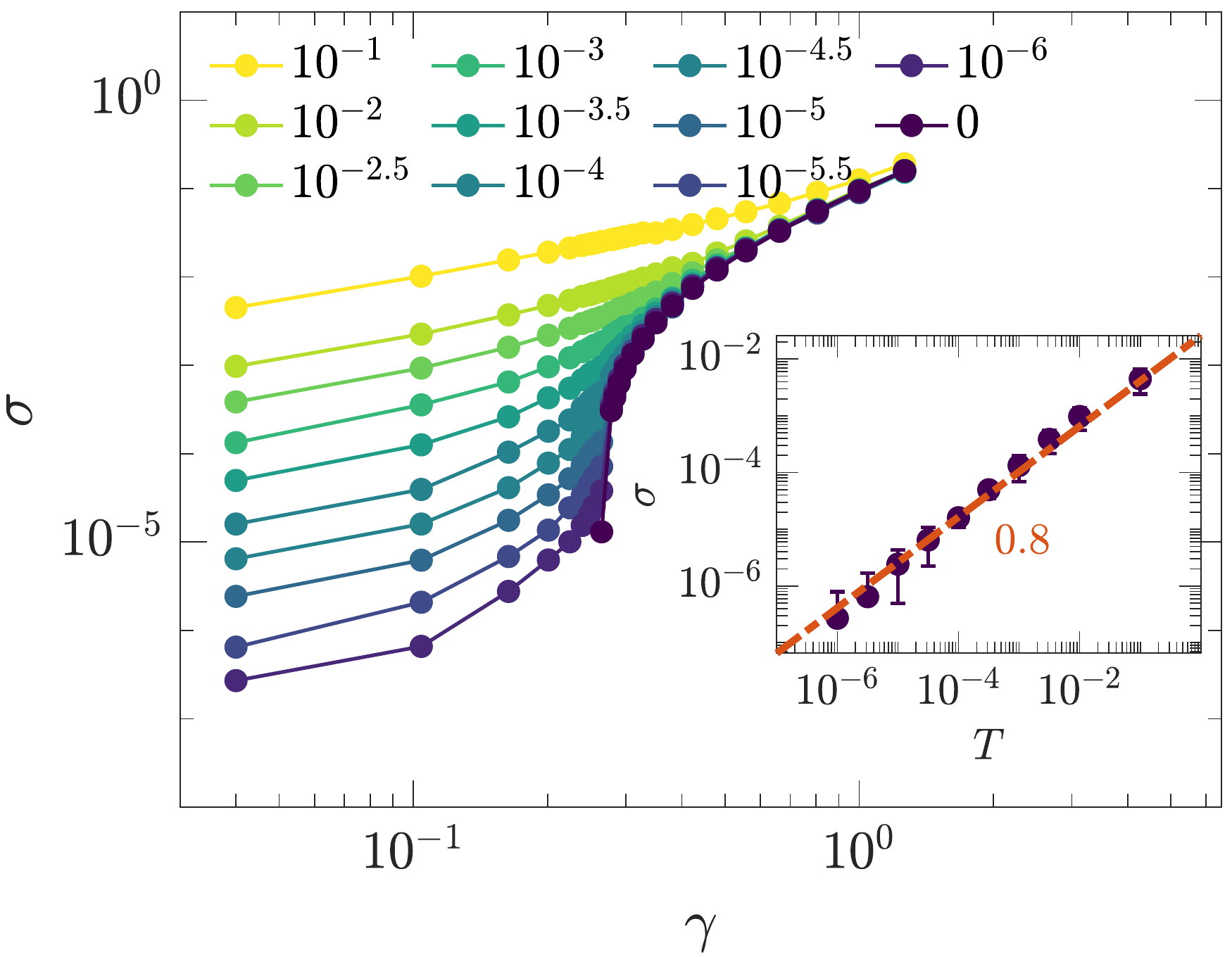}
	\caption{\label{fig_sup_mc_stress_tr}Shear stress versus strain for diluted triangular networks at $z=3.3$ and at various temperature. The inset shows the scaling behavior of shear stress versus temperature in the linear regime $\gamma = 0.04 < \gamma_c$.}
\end{figure}

The shear stress $\sigma$ versus shear strain $\gamma$ behavior for randomly diluted triangular networks at $z=3.3$ is presented in Figure \ref{fig_sup_mc_stress_tr}. At zero temperature, the network is floppy and does not exhibit any shear stress until it reaches the critical strain $\gamma_c$. As the temperature increases, the network becomes stable and a finite shear stress emerges below the critical strain $\gamma_c$. In the linear regime $\gamma<\gamma_c$, we find an anomalous temperature dependence $\sigma \sim T^{0.8}$ (see the inset).

\begin{figure}[!h]
	\centering
	\includegraphics[width=5cm, height=5cm,keepaspectratio]{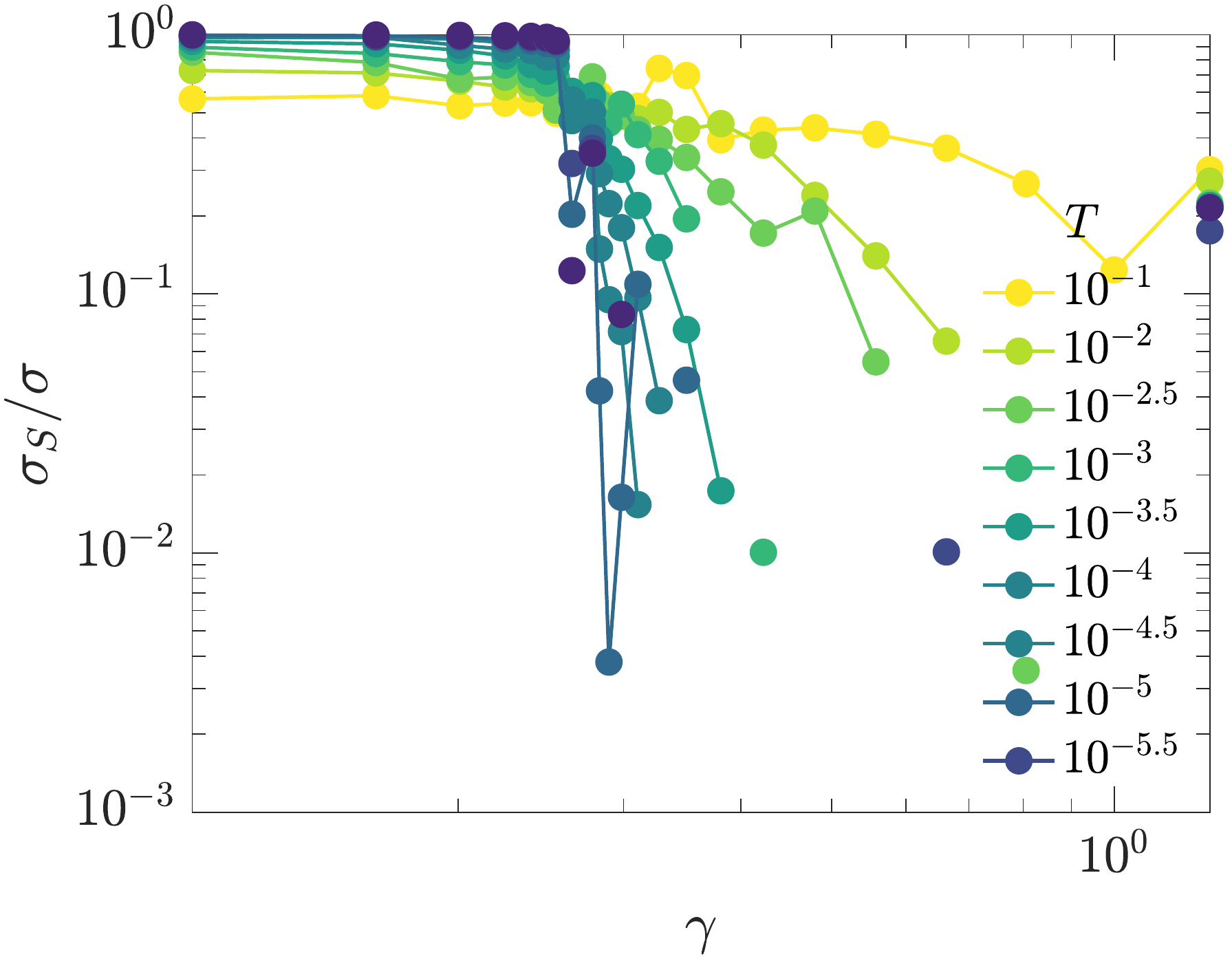}
	\caption{\label{fig_sup_mc_sigma_s_over_sigma} The ratio of entropic stress to overall stress in diluted triangular networks at $z=3.3$.}
\end{figure}

The total shear stress $\sigma$ can be decomposed into energetic and entropic contributions by differentiating the free energy of the system $F = E - TS$ with respect to the shear strain $\gamma$ \cite{dennison_fluctuation-stabilized_2013}
\begin{equation}\label{eq_sup_stress_contributions}
	\sigma = \frac{1}{V} \frac{\partial F}{\partial \gamma} = \frac{1}{V} \frac{\partial E}{\partial \gamma} - \frac{T}{V} \frac{\partial S}{\partial \gamma},
\end{equation}
where $V$, $E$, $T$, and $S$ denote the volume (area), energy, temperature, and entropy of the system, respectively. The energetic stress, represented by $\sigma_{E} = \frac{1}{V} \frac{\partial E}{\partial \gamma}$, and the entropic stress, represented by $\sigma_{S} = - \frac{T}{V}\frac{\partial S}{\partial \gamma}$, can thus be obtained. In our Monte Carlo simulations, we can easily compute the ensemble average of the energy, which provides direct access to the energetic stress $\sigma_{E}$. The entropic stress can then be computed by subtracting the energetic contributions from the ensemble-averaged total stress, i.e., $\sigma_{S} = \sigma - \sigma_{E}$ \cite{dennison_fluctuation-stabilized_2013}. The entropic part of shear stress $\sigma_S$ dominates the response in the small strain regime (as shown in Fig. \ref{fig_sup_mc_sigma_s_over_sigma}). However, at the critical strain,$\sigma_S$ decreases significantly and eventually disappears in the large strain regime because the system becomes predominantly stretching-dominated.

\renewcommand\thefigure{C\arabic{figure}} 
\setcounter{figure}{0}
\renewcommand\theequation{C\arabic{equation}}
\setcounter{equation}{0}

\section{Honeycomb lattice model results}
\begin{figure}[!h]
	\centering
	\includegraphics[width=5cm, height=5cm,keepaspectratio]{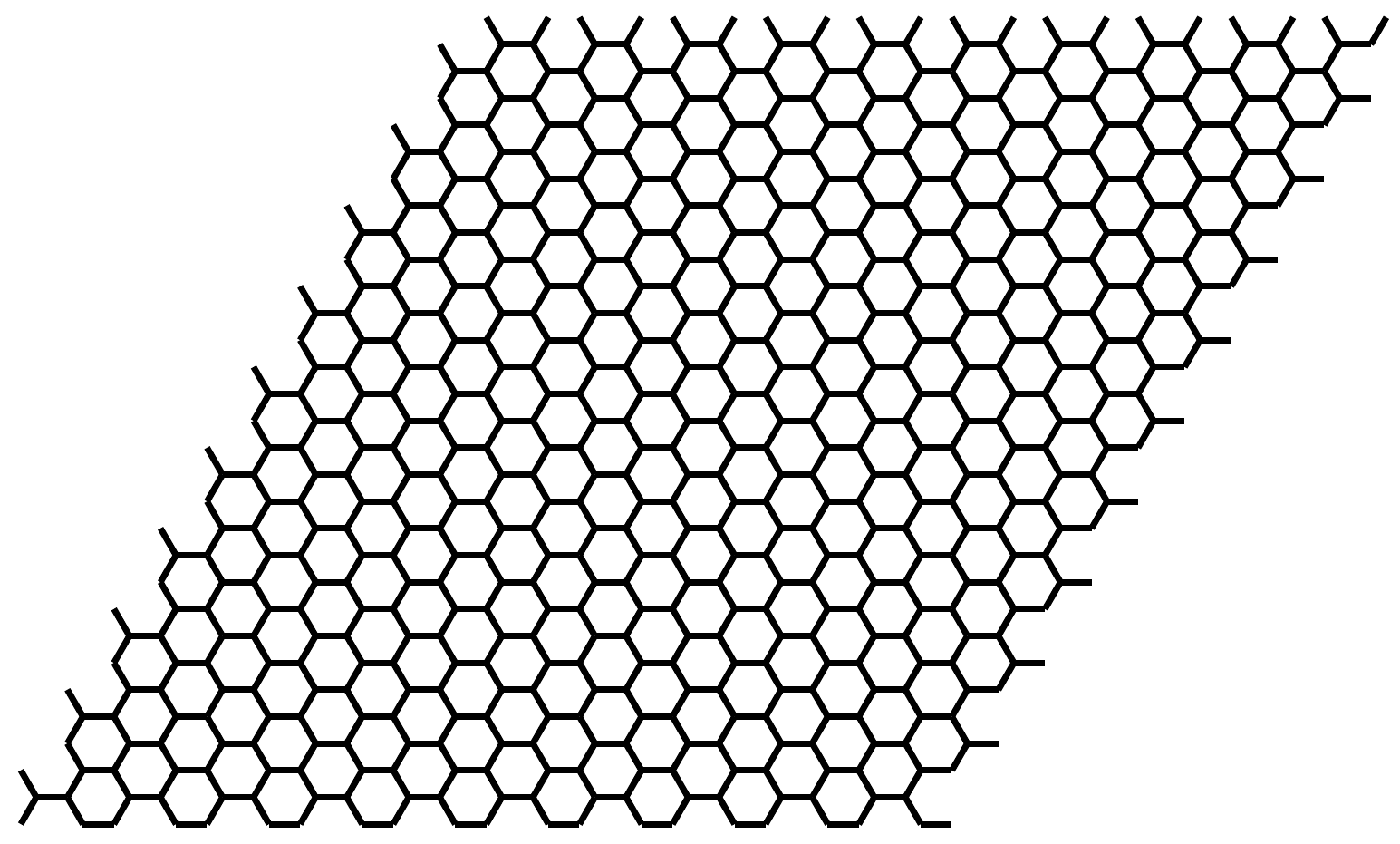}
	\caption{\label{fig_sup_mc_network_hc} A regular honeycomb lattice network. Note that this is the undeformed state of our model. We use a parallelogram to simulate our network due to its simplicity.}
\end{figure}

\begin{figure}[!h]
	\centering
	\includegraphics[width=8cm, height=8cm,keepaspectratio]{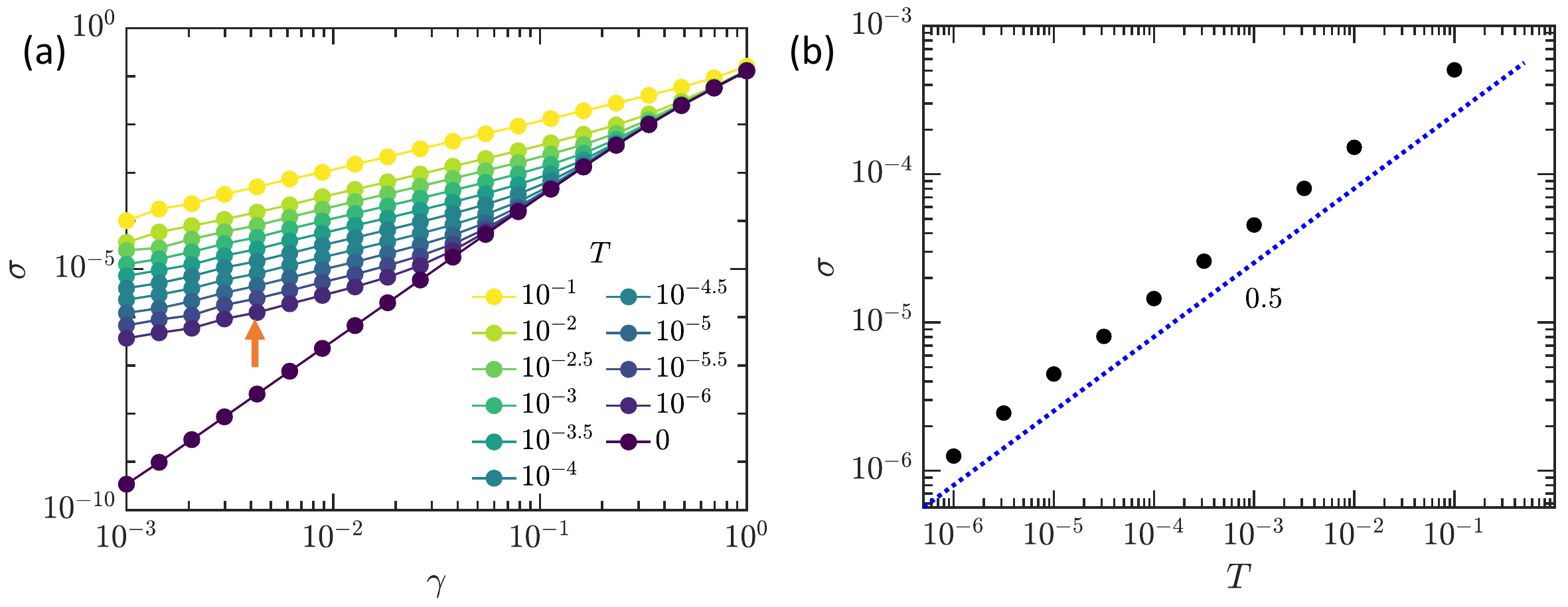}
	\caption{\label{fig_sup_mc_sigma_hc} (a) The shear stress versus strain for a regular honeycomb lattice with no distortion at various $T$ values. (b) Scaling behavior of shear stress data in (a) in the small strain regime (the arrow in (a) shows the strain level) versus $T$. The data are obtained for a lateral size of $W=90$ and averaged over $10^8$ MC steps.}
\end{figure}

\begin{figure}[!h]
	\centering
	\includegraphics[width=8cm, height=8cm,keepaspectratio]{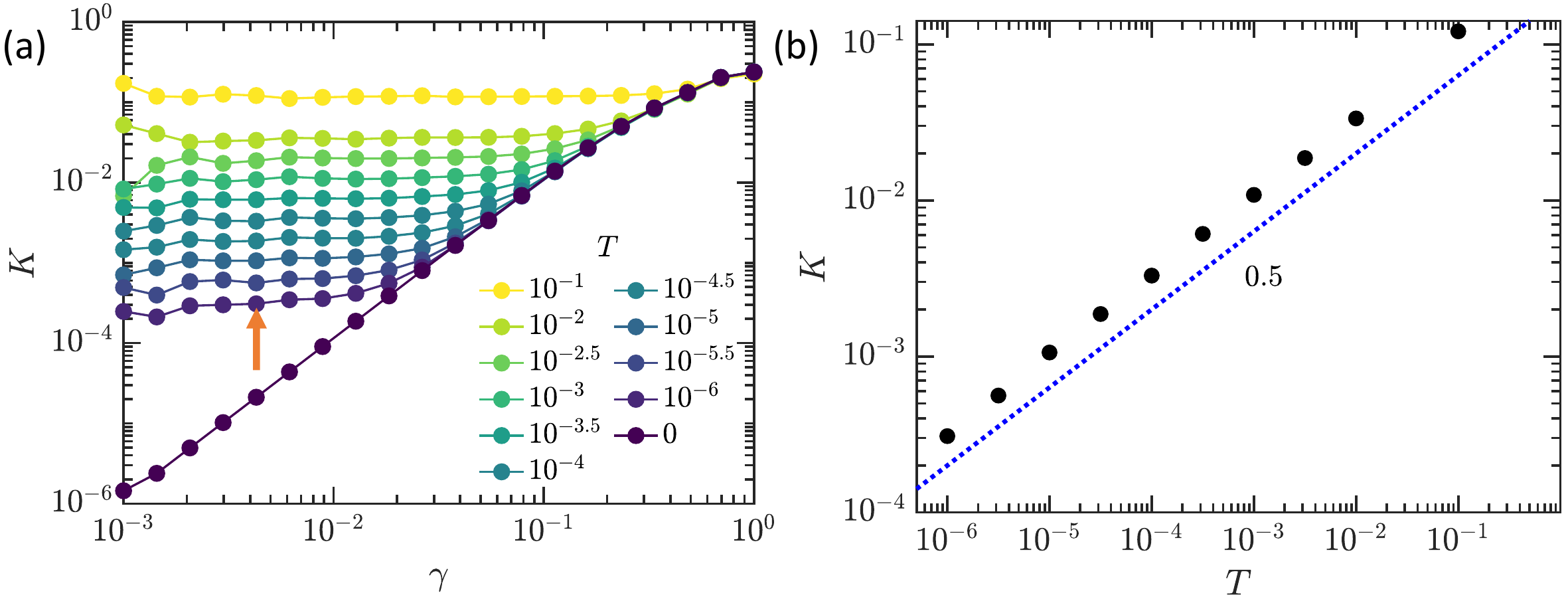}
	\caption{\label{fig_sup_mc_K_hc} (a) The differential shear modulus versus strain for a regular honeycomb lattice with no distortion at various $T$ values. (b) Scaling behavior of $K$ data in (a) in the small strain regime (the arrow in (a) shows the strain level) versus $T$. The data are obtained for a lateral size of $W=90$ and averaged over $10^8$ MC steps.}
\end{figure}

The regular honeycomb (hexagonal) lattice is an ideal option for studying the anomalous temperature-dependent behavior of the shear modulus. This is because the lattice's symmetry ensures its stability under any finite strain \cite{rens_nonlinear_2016}. Figure \ref{fig_sup_mc_network_hc} shows an example of a small honeycomb structure at its undeformed state. At $T=0$, the shear stress of this model exhibits a finite value in the small strain regime, as expected (see Fig. \ref{fig_sup_mc_sigma_hc}a). As temperature increases, the shear stress displays a power law scaling relationship with $T$, characterized by an exponent of 0.5 (as depicted in Fig. \ref{fig_sup_mc_sigma_hc}b). This behavior is also evident from the analysis of the differential shear modulus, as shown in Figure \ref{fig_sup_mc_K_hc}.

\begin{thebibliography}{56}
\bibitem [{Hea()}]{HeatCapacity}%
	\BibitemOpen
	\href@noop {} {}\bibinfo {note} {Similarly, direct thermal signatures such as
		the heat capacity may not exhibit singular behavior that is apparent, e.g.,
		when studying quantities corresponding to derivatives of the free energy with
		respect to shear.}\BibitemShut {Stop}%
\end{thebibliography}

%

\end{document}